\documentclass[seceqn]{elsart}

\usepackage{amsmath}
\usepackage{amsfonts}
\usepackage{amssymb}

\makeatletter
\newif\if@noprintpreprintstamp

\@noprintpreprintstamptrue  

\if@noprintpreprintstamp
\def\ps@copyright{}
\else
\fi
\makeatother

\begin{document}

\renewcommand{\r}{\mathbf{r}}
\newcommand{\anglesol}{\tilde{\Omega}}
\newcommand{\dist}{\vartheta}
\newcommand{\mynormalordering}[1]{::#1\-::\,}
\newcommand{\exc}{{\mathrm{exc}}}
\renewcommand{\atop}[2]{\genfrac{}{}{0pt}{}{#1}{#2}}

\begin{frontmatter}

\title{Debye--H\"uckel theory for two-dimensional Coulomb systems
living on a finite surface without boundaries}

\author{Gabriel T\'ellez}

\address{Departamento de F\'{\i}sica\\
Universidad de Los Andes\\
A.A. 4976\\
Bogot\'a, Colombia\\
}

\ead{
gtellez@uniandes.edu.co
}

\begin{abstract}
We study the statistical mechanics of a multicomponent two-dimensional
Coulomb gas which lives on a finite surface without boundaries. We
formulate the Debye--H\"uckel theory for such systems, which describes
the low-coupling regime. There are several problems, which we address,
to properly formulate the Debye--H\"uckel theory. These problems are
related to the fact that the electric potential of a single charge
cannot be defined on a finite surface without boundaries. One can only
define properly the Coulomb potential created by a globally neutral
system of charges. As an application of our formulation, we study, in
the Debye--H\"uckel regime, the thermodynamics of a Coulomb gas living
on a sphere of radius $R$. We find, in this example, that the grand
potential (times the inverse temperature) has a universal finite-size
correction $(1/3) \ln R$. We show that this result is more general:
for any arbitrary finite geometry without boundaries, the grand
potential has a finite-size correction $(\chi/6) \ln R$, with $\chi$
the Euler characteristic of the surface and $R^2$ its area.
\end{abstract}


\maketitle

\end{frontmatter}

\section{Motivation}

In this paper, we study two-dimensional Coulomb systems, for instance
plasmas or electrolytes, which live on a finite surface without
boundaries. The simplest example of such geometry is the sphere. This
kind of geometry has been used in numerical simulations of charged
systems~\cite{Caillol-Levesque-Weis-Hansen-JSP,Caillol-Levesque-PRB,Caillol-Levesque-JCP,Caillol-JCP-hyperspherical}
as an alternative to the Ewald method, since it avoids the problem of
the boundary conditions. There has been also several theoretical
studies of Coulomb systems on the sphere, in particular, the
statistical mechanics of both the one-component plasma and the charge
symmetric two-component plasma, can be exactly solved for a special
value of the Coulomb
coupling~\cite{Caillol-sphere-2D,dosdesfera,Tellez-Forrester-2dOCP-Gamma=4-6}.
The purpose of this work is to formulate the Debye--H\"uckel theory
for a generic multi-component Coulomb system living on a surface
without boundaries. The Debye--H\"uckel theory describes the
low-coupling regime of the system.

There are several difficulties which make this problem
non-trivial. First, it is not possible to define the Coulomb potential
as the inverse of the Laplacian operator, since on a surface without
boundaries, the Laplacian has no inverse. Physically, this means that
the electric potential of a single charge, on the sphere, does not
exist. One can only properly define the electric potential created by
a globally neutral configuration of charges. Thus, the partition
function of such system should be restricted to neutral
configurations. This problem is not present on an infinite surface, or
a surface with boundaries, where any excess charge can go to infinity
or to the boundaries, and does not affect the bulk thermodynamics of
the system.

There are several equivalent ways to formulate the Debye--H\"uckel
theory. Recently, the author and a collaborator proposed one, which is
particularly appropriate to study confined Coulomb
systems~\cite{finite-size-DH,general-DH}. This method is based on the
Hubbard--Stratonovich transformation, also known as the sine-Gordon
transformation, when applied to Coulomb systems. However, for a
Coulomb system living in a surface without boundary, this method is
not directly applicable, because: 1) to use the sine-Gordon
transformation one should consider all possible configurations of the
system, including the globally charged ones, 2) the Laplacian, which
in a flat geometry appears as the inverse of the Coulomb potential,
has no inverse on a surface without boundaries.

The outline of this work is as follows. In the next section, we will
show how to overcome the difficulties mentioned above and how the
approach of Refs.~\cite{finite-size-DH,general-DH} can be extended to
geometries without boundaries. For simplicity we will consider the
case of the sphere, however the method presented in
Sec.~\ref{sec:sine-Gordon-no-boundaries} is general enough to be
applicable to any surface without boundaries. In
Sec.~\ref{sec:sphere-DH}, using the results of
Sec.~\ref{sec:sine-Gordon-no-boundaries}, we will explicitly compute
the grand potential and other thermodynamics functions of a Coulomb
system living on a sphere in the Debye--H\"uckel regime. We will also
compute the finite-size expansion of the grand potential, and show the
existence of a finite-size correction $(1/3)\ln R$ to the grand
potential (times the reduced inverse temperature), where $R$ is the
radius of the sphere. Finally in Sec.~\ref{sec:finite-size}, with the aid
of some results from Ref.~\cite{general-DH}, we will show that for a
general geometry without boundaries, the grand potential has a
logarithmic finite-size correction $(\chi/6) \ln R$ where $\chi$ is
the Euler characteristic of the manifold where the system lives and
$R^2$ is proportional to the area of the manifold. 


\section{Modified sine-Gordon transformation}
\label{sec:sine-Gordon-no-boundaries}

\subsection{Coulomb potential and energy of a pair of pseudocharges}

The Coulomb potential is usually defined as the solution of Poisson
equation
\begin{equation}
  \label{eq:Poisson}
  \Delta v(\r,\r')=-2\pi \delta(\r,\r')
\end{equation}
with appropriate boundary conditions. For a flat plane geometry, the
potential is
\begin{equation}
  \label{eq:v0}
  v^{0}(\r,\r')=-\ln\frac{|\r-\r'|}{L}
\end{equation}
which satisfies~(\ref{eq:Poisson}) and the boundary condition $\nabla
v^{0}(\r,\r')\to 0$ as $|\r-\r'|\to\infty$. In equation~(\ref{eq:v0}), $L$
is an arbitrary constant which fixes the zero of the electric
potential.

On a sphere, of radius $R$, and more generally on any finite surface
without boundaries, it is not possible to define the Coulomb potential
from Poisson equation~(\ref{eq:Poisson}). This equation has no
solution on the sphere: the electric potential of a single charge
cannot be defined. Instead, following~\cite{Caillol-Levesque-JCP}, we
should consider that the system is composed of pseudocharges. A
pseudocharge is the combination a unit point charge and a uniform
background of opposite charge spread over the sphere. The electric
potential $v$ created by a unit pseudocharge satisfies
\begin{equation}
  \label{eq:Poisson-sphere}
  \Delta v(\anglesol,\anglesol') = -2\pi \left(
  \delta(\anglesol,\anglesol')-\frac{1}{4\pi R^2}
  \right)
  \,.
\end{equation}
with $\anglesol=(\theta,\varphi)$ the spherical coordinates of source and
$\anglesol'=(\theta',\varphi')$ the location where the potential is
computed, and $\delta(\anglesol,\anglesol')=
\delta(\cos\theta-\cos\theta')\delta(\varphi-\varphi')/R^2$ is the Dirac
distribution on the sphere of radius $R$.

Decomposing in spherical harmonics $Y_{\ell m}$, one can write down
the solution of the previous equation as
\begin{equation}
  \label{eq:Coulomb-harmonics}
  v(\anglesol;\anglesol')=2\pi\sum_{\ell=1}^{\infty
  }\sum_{m=-\ell}^{\ell} \frac{1}{\ell(\ell+1)}\overline{Y_{\ell
  m}(\theta ^{\prime },\varphi ^{\prime })}Y_{\ell m}(\theta ,\varphi
  ) + V_0
\end{equation}
with $V_0$ an arbitrary constant. Now, using~\cite{grad}
\begin{equation}
  -\ln \sin \frac{\dist}{2}
     =\sum_{\ell=1}^{\infty }
  \frac{2\ell+1}{2\ell(\ell+1)}\,
  P_{\ell}(\cos \dist )+\frac{1}{2}
\end{equation}
where $P_{\ell}(x)$ are the Legendre polynomials of order $\ell$, and
$\dist$ is the angle between $\anglesol$ and $\anglesol'$, we can
write the Coulomb potential~(\ref{eq:Coulomb-harmonics}) as
\begin{equation}
  \label{eq:Coulomb-pot-sphere}
  v(\theta,\varphi;\theta',\varphi) 
  =-\ln \sin
  \frac{\dist}{2} -\frac{1}{2} + V_0
  \,.
\end{equation}
The potential energy $E_{ij}$ of two pseudocharges $q_i$
and $q_j$ located at $\anglesol_i$ and $\anglesol_j$ can be decomposed
as $E_{ij}=V_{ij}+\varepsilon_i+\varepsilon_j$, with $V_{ij}$ the
interaction energy and $\varepsilon_i$ ($\varepsilon_j$) the
self-energy of the pseudocharge $q_i$ ($q_j$). The interaction energy
is
\begin{equation}
  V_{ij}=q_{i} \int v(\anglesol_i,\anglesol)\rho_{j}(\anglesol)\,R^2 d\anglesol
\end{equation}
where
$\rho_{j}(\anglesol)=q_{j}\delta(\anglesol,\anglesol_j)-q{_j}/(4\pi
R^2)$ is the charge density of the pseudocharge $q_j$. We find
\begin{equation}
  V_{ij}=-q_i q_j \left( \ln\sin\frac{\dist_{ij}}{2}-\frac{1}{2}
  \right)
\end{equation}
with $\dist_{ij}$ the angle between $\anglesol_i$ and $\anglesol_j$
measured from the center of the sphere. Notice that the interaction
energy does not depend on the arbitrary constant $V_0$.

Following~\cite{Caillol-Levesque-JCP}, the self-energy $\varepsilon_i$
of a pseudocharge can be computed by integrating the square of the
electric field $\mathbf{E}(\anglesol)=-\nabla
v(\anglesol,\anglesol_i)$ created by the pseudocharge on the whole
sphere. However this self-energy is infinite for a point
charge. Instead we can replace, initially, the point charge by a small
disk of radius $a$ with the charge $q_i$ spread over its perimeter and
we compute
\begin{equation}
  \label{eq:self-pseudo-pre}
  \varepsilon_i=\frac{R^2}{4\pi}\int |\mathbf{E}(\anglesol)|^2
  \,d\anglesol
  - \varepsilon^{s}_{i}
  \,.
\end{equation}
We have subtracted $\varepsilon_{i}^{s}$, the self-energy of the
charge $q_i$ alone, without its neutralizing background. This
self-energy $\varepsilon_{i}^{s}$ is not properly defined on the
sphere, since it correspond to a non-neutral charge configuration. As
in Ref.~\cite{Caillol-Levesque-JCP} we adopt the prescription of
taking $\varepsilon_{i}^{s}$ as the self-energy of the charge $q_i$ on
a flat surface
\begin{equation}
  \label{eq:prescript-self-s}
  \varepsilon_{i}^{s}=-\frac{q_i^2}{2}\ln \frac{a}{L}
\end{equation}
In the limit of a point charge, $a\to0$, the self-energy of the
pseudocharge~(\ref{eq:self-pseudo-pre}) is finite
\begin{equation}
  \label{eq:self-pseudo}
  \varepsilon_i=
  \frac{q_i^2}{4}\left(-1+2\ln\frac{2R}{L}\right)
  \,.
\end{equation}

Let us now study a Coulomb system on the sphere composed of several
species of pseudocharges $q_{\alpha}$, $\alpha=1,\cdots,s$. Each
species $\alpha$ has $N_{\alpha}$ pseudocharges. The position of the
$i$-th pseudocharge of the species $\alpha$ will be denoted by
$\anglesol_{\alpha,i}$. The total potential energy of the system is
\begin{equation}
  \label{eq:hamilt}
  H=\frac{1}{2} \sum_{\alpha,\gamma}
  \sum_{i=1}^{N_\alpha}\sum_{j=1}^{N_\gamma} {}^\prime q_\alpha q_\gamma
  v_c(\anglesol_{\alpha,i},\anglesol_{\gamma,j})+\sum_{\alpha}
  \sum_{i=1}^{N}
  \varepsilon_{\alpha}
\end{equation}
with
\begin{equation}
  \label{eq:vc}
  v_c(\anglesol_{\alpha,i},\anglesol_{\gamma,j})=-\ln\sin(\dist_{ij}/2)-1/2
\end{equation}
and $\dist_{ij}$ the angle from the center of the sphere between
$\anglesol_{\alpha,i}$ and $\anglesol_{\gamma,j}$.  The prime in the
summation in~(\ref{eq:hamilt}) means that the term when
$\alpha=\gamma$ and $i=j$ must be omitted.

We shall call ``neutral'' any configuration of pseudocharges
satisfying $\sum_{\alpha} q_{\alpha} N_{\alpha}=0$. For a neutral
configuration, the backgrounds of the pseudocharges cancel each other,
and the system is equivalent to a system of charged particles without
neutralizing backgrounds. For those neutral configurations, we have
$(\sum_{\alpha}\sum_{i=1}^{N_{\alpha}} q_{\alpha}
)^2=0=\sum_{\alpha}\sum_{i=1}^{N_{\alpha}}
q_{\alpha}^2+\sum_{\alpha,\gamma}\sum_{i=1}^{N_{\alpha}}
\sum_{j=1}^{N_{\gamma}}{}'q_{\alpha}q_{\gamma}$.
Then, the potential energy can be put in the form
\begin{equation}
  H_{\mathrm{neutral}}=-\frac{1}{2}\sum_{\alpha,\gamma}\sum_{i,j}^{}
  {}^\prime q_{\alpha} q_{\gamma}
  \ln\left(\frac{2R}{L}\sin\frac{\dist_{ij}}{2}\right) \,.
\end{equation}
In the flat limit $R\to\infty$, keeping $r_{ij}=R\dist_{ij}$ finite,
the above expression reduces to $H_{\mathrm{flat}}=
-(1/2)\sum_{\alpha,\gamma}\sum_{i,j}^{\prime} q_{\alpha} q_{\gamma}
\ln(r_{ij}/L)$, which is the hamiltonian of a Coulomb system in the
flat geometry. Thus, the prescription~(\ref{eq:self-pseudo}) for the
self-energy is a reasonable one, if one wishes to recover the results
for the flat geometry in the limit
$R\to\infty$~\cite{Caillol-JCP-hyperspherical}. Notice, however, that
the thermodynamics of the Coulomb system on the flat geometry can only
be recovered if we consider only neutral configurations from the start
on the sphere.

\subsection{Sine-Gordon transformation}

We shall work in the grand canonical ensemble. The fugacity of the
species $\alpha$ will be denoted by $\zeta_{\alpha}$. We define as
usual $\beta=1/(k_B T)$, were $T$ is the absolute temperature and
$k_B$ is the Boltzmann constant. The grand partition function reads
\begin{equation}
  \label{eq:part-funct-1}
  \Xi=\sum_{N_{1}=0}^{\infty } \cdots \sum_{N_{s}=0}^{\infty }
  \delta_{\sum_{\alpha}q_\alpha N_{\alpha},0}\, \frac{ \zeta
  _{1}^{N_{1}}\cdots\zeta _{r}^{N_{s}}}{N_{1}!\ldots N_{s}!}
  \int\cdots\int
  e^{-\beta H}\prod_{\alpha =1}^{r}\prod_{i=1}^{N_{\alpha }}R^2d
  \anglesol_{\alpha,i}
\end{equation}
The Kronecker symbol $\delta_{\sum_{\alpha}q_\alpha N_{\alpha},0}$
ensures that we only take neutral configurations.

Introducing the charge density $\rho(\anglesol)=\sum_{\alpha,i}
q_{\alpha}\delta(\anglesol-\anglesol_{\alpha,i})$, the
hamiltonian~(\ref{eq:hamilt}) can be put in the form
\begin{equation}
  H=\frac{1}{2} \int \rho(\anglesol)\rho(\anglesol')
v_c(\anglesol,\anglesol') \, R^2 d\anglesol \,R^2 d\anglesol'
+\sum_{\alpha,i} \left[\varepsilon_{\alpha} - \frac{q_{\alpha}^2}{2}
v_c(\anglesol_{\alpha,i},\anglesol_{\alpha,i})\right]
\end{equation}
The term $v_c(\anglesol_{\alpha,i},\anglesol_{\alpha,i})$ which is
subtracted corresponds to the term when, for the same particle,
$\anglesol=\anglesol'$ in the integral which should not be included in
the hamiltonian. Using the explicit expression of the self
energy~(\ref{eq:self-pseudo}), we notice that $\varepsilon_{\alpha} -
q_{\alpha}^2 v_c(\anglesol_{\alpha,i},\anglesol_{\alpha,i})/2= -
q_{\alpha}^2 v^{0}(\r_{\alpha,i},\r_{\alpha,i})/2$ is the self-energy
of a particle in the flat geometry.

Let us decompose the hamiltonian in spherical harmonics
$Y_{\ell m}$. Let
\begin{equation}
  \rho(\anglesol)=\sum_{\ell=0}^{\infty}\sum_{m=-\ell}^{\ell}
 \rho_{\ell m} Y_{\ell m}(\anglesol) \,.
\end{equation}
Using~(\ref{eq:Coulomb-harmonics}) we have
\begin{equation}
  H=\pi R^4 \sum_{\ell=1}^{\infty} \sum_{m=-\ell}^{\ell}
  \frac{|\rho_{\ell m}|^2}{\ell(\ell+1)}
  -\sum_{\alpha,i} q_{\alpha}^2
  v^{0}(\r_{\alpha,i},\r_{\alpha,i})/2
\,.
\end{equation}

There are two problems to proceed to apply the sine-Gordon
transformation to the grand partition function. First, the potential
$v_c$ from equation~(\ref{eq:vc}), has no inverse. This can be seen in
the expansion in spherical harmonics given by
equation~(\ref{eq:Coulomb-harmonics}), with the term for $\ell=0$,
$V_0=0$. Second, to apply the sine-Gordon transformation one needs to
consider all configurations of the system including the non-neutral
ones. However, both problems are related and can be solved jointly.

Let us write $\delta_{\sum_{\alpha}q_\alpha
  N_{\alpha},0}=\lim_{\epsilon\to\infty}\exp\left[-\beta \epsilon
  \left(\sum_{\alpha} q_{\alpha} N_{\alpha}\right)^2/4\right]$ in the
  partition function~(\ref{eq:part-funct-1}). Then we have
  $\Xi=\lim_{\epsilon\to\infty}\Xi_{\epsilon}$, where $\Xi_{\epsilon}$
  is a partition function not restricted to neutral configurations and
  with hamiltonian
  \begin{equation}
    H_{\epsilon}=H+\epsilon\left[\sum_{\alpha} q_{\alpha}
    N_{\alpha}\right]^2/4
    \,.
  \end{equation}
  This last term can be seen as a $\ell=0$ component to the
  interaction between pairs of particles. Indeed, since
  $\rho_{00}=R^{-2} \sum_{\alpha,i} q_{\alpha}
  Y_{00}(\Omega_{\alpha,i}) = (4\pi)^{-1/2} R^{-2} \sum_{\alpha}
  q_{\alpha} N_{\alpha}$, we have
  \begin{equation}
    \label{eq:hamilt-epsilon}
    H_{\epsilon}=\pi R^4 \left[\rho_{00}^2 \epsilon+
      \sum_{\ell=1}^{\infty} \sum_{m=-\ell}^{\ell}
      \frac{|\rho_{\ell,m}|^2}{\ell(\ell+1)}\right] -\sum_{\alpha,i}
    q_{\alpha}^2 v^{0}(\r_{\alpha,i},\r_{\alpha,i})/2 \,.
  \end{equation}
This solves also the second problem we faced before to apply the
sine-Gordon transformation: the quadratic form in the hamiltonian has
now a non-vanishing $\ell=0$ term and it is now invertible. At the end
of the calculations we should take $\epsilon\to\infty$. Thus, the
$\ell=0$ term of~(\ref{eq:hamilt-epsilon}) goes to infinity. If we
remember that the Coulomb potential is the inverse of the Laplacian
operator, this is a reminder that $1/[\ell(\ell+1)]$ diverges when
$\ell=0$. This $\ell=0$ term that we naturally introduce to take into
account only the neutral configurations, has also modified the pair
potential to render it invertible. Its inverse is the Laplacian, plus
a $\ell=0$ component proportional to $1/\epsilon$.

We now proceed as usual and perform the sine-Gordon transformation. In
Refs.~\cite{finite-size-DH,general-DH} we showed that the role of the
flat self-energy term $v^{0}$ in~(\ref{eq:hamilt-epsilon}) is to
regularize the ultraviolet divergence of the other terms. Therefore we
will concentrate our efforts only in the part $H_{\epsilon}'=\pi R^4
\left[\rho_{00}^2 \epsilon+ \sum_{\ell=1}^{\infty}
\sum_{m=-\ell}^{\ell}
|\rho_{\ell,m}|^2/[\ell(\ell+1)]\right]$. Performing the sine-Gordon
transformation~\cite{Samuel-sG}, we have
\begin{equation}
  \label{eq:functional-int}
  \Xi_{\epsilon}=\frac{Z}{Z_0}=\frac{
  \int \prod_{\ell=0}^{\infty}\prod_{m=-\ell}^{\ell}
    d\phi_{\ell m}
  \exp[-S(\{\zeta_{\alpha}\})]}{
  \int \prod_{\ell=0}^{\infty}\prod_{m=-\ell}^{\ell}
  d\phi_{\ell m}
  \exp[-S(0)]}
\end{equation}
with the action 
\begin{eqnarray}
  S(\{\zeta_{\alpha}\})&=& -\frac{1}{2}\sum_{\ell,m} A_{\ell
  m}\phi_{\ell m}^2 
  \nonumber\\
  &&- \sum_{\alpha} \zeta_{\alpha} \int
  \mynormalordering{ \exp\left[-i\beta
  q_{\alpha}\sum_{\ell=0}^{\infty}\sum_{m=-\ell}^{\ell} \phi_{\ell m}
  Y_{\ell m}(\anglesol)\right]}\,R^2 d\anglesol  .
\end{eqnarray}
We have defined $A_{\ell m}=\beta \ell(\ell+1)/(2\pi)$ if $\ell\geq 1$
and $A_{00}=\beta \epsilon/(2\pi)$. $(A_{\ell m})$ represents an
operator which is the inverse of the modified Coulomb potential. It is
proportional to the Laplacian plus a $\ell=0$
component. $\phi(\anglesol)=\sum_{\ell,m} \phi_{\ell m} Y_{\ell
  m}(\anglesol)$ is the traditional auxiliary field introduced in the
sine-Gordon transformation. The notation $\mynormalordering{\cdots}$
is defined as
\begin{equation}
  \mynormalordering{
    e^{-i\beta q_{\alpha} \phi(\anglesol)}
  }
  =    e^{-i\beta q_{\alpha} \phi(\anglesol)}
  e^{\beta q_{\alpha}^2 v^{0}(\r,\r)}
\end{equation}
with $\r=(R,\anglesol)$, and can be understood as a sort of normal
ordering operator, for details see~\cite{3Dslab}.

The $\ell=0$ term in the action $S$ is special, because at the end we
will be interested in the limit $\epsilon\to\infty$. Let us do the
change of variable in the functional integral $X=\sqrt{\beta/(2\pi
\epsilon)}\phi_{00}$. The action is now
\begin{eqnarray}
  \label{eq:action-S}
  S(\{\zeta_{\alpha}\})&=&
  \sum_{\ell\geq  1,\,m} \frac{\beta\ell(\ell+1)}{4\pi}
  \phi_{\ell m}
  +\frac{X^2}{2}
  \\
  &&
  -\sum_{\alpha}
  \zeta_{\alpha}
  \int
  \mynormalordering{
    e^{-i\beta
      q_{\alpha}\left[\sum_{\ell\geq1,\,m}\phi_{\ell m}
      Y_{\ell m}(\anglesol)
      +\sqrt{\epsilon/(2\beta)} X \right]
  }}\,R^2 d\anglesol
  \,.
  \nonumber
\end{eqnarray}

\subsection{Steepest descent}

We now consider the low-coupling or Debye--H\"uckel regime, when
$\beta e^2\ll 1$, with $e$ the elementary charge (all charges
$q_{\alpha}=z_{\alpha} e$ are supposed to be multiples of $e$,
$z_{\alpha}\in\mathbb{Z})$. In the low-coupling approximation, the
action~(\ref{eq:action-S}) is expanded to the quadratic order around
the solution of the stationary action equation $\delta S/\delta
\phi=0$, then the functional integral in
equation~(\ref{eq:functional-int}) is Gaussian and can be performed
exactly. The stationary action equation $\delta S/\delta \phi=0$
reads, for $\ell\geq 1$,
\begin{equation}
  \label{eq:minim-lm}
  \frac{\beta}{2\pi} \ell(\ell+1)\phi_{\ell m}
  +\sum_{\alpha} i\beta q_{\alpha}\zeta_{\alpha}
  \int Y_{\ell m}(\anglesol)
   e^{-i\beta
      q_{\alpha}\left[\sum_{\ell\geq1,\,m}\phi_{\ell m}
      Y_{\ell m}(\anglesol)
      +\sqrt{\epsilon/(2\beta)} X \right]
   } R^2 d\anglesol = 0
\end{equation}
and, for $\ell=0$,
\begin{equation}
  \label{eq:minim-00}
  X
  +\sum_{\alpha} i\beta q_{\alpha}\zeta_{\alpha}
  \sqrt{\frac{\epsilon}{2\beta}}
  \int 
  e^{-i\beta
    q_{\alpha}\left[\sum_{\ell\geq1,\,m}\phi_{\ell m}Y_{\ell m}(\anglesol)
    +\sqrt{\epsilon/(2\beta)} X \right]
  } R^2 d\anglesol = 0
\end{equation}
Since the system is homogeneous and isotropic, we look for a constant
solution, $\phi_{\ell m}=0$ for $\ell\geq 1$, which naturally
satisfies~(\ref{eq:minim-lm}). Then equation~(\ref{eq:minim-00})
reduces to
\begin{equation}
  \label{eq:minim-exc}
  \frac{X}{2\pi\sqrt{2\beta \epsilon}}
  +iR^2 \sum_{\alpha}  q_{\alpha}\zeta_{\alpha}
  e^{-i\beta
    q_{\alpha}\sqrt{\epsilon/(2\beta)} X }
  =0
\end{equation}
Since we are interested in the limit $\epsilon\to\infty$, the first
term can be neglected and we have
\begin{equation}
  \label{eq:minim-app}
  \sum_{\alpha}  q_{\alpha}\zeta_{\alpha}
  e^{-i\beta
    q_{\alpha}\sqrt{\epsilon/(2\beta)} X }
  =0
\end{equation}
Notice that, if the fugacities are chosen to satisfy the
pseudoneutrality condition $\sum_{\alpha}q_{\alpha}\zeta_{\alpha}=0$,
then $X=0$ is a solution of~(\ref{eq:minim-app}). But $X=X_n=n 2\pi
\sqrt{2/(\beta e^2\epsilon)}$ is also a solution with
$n\in\mathbb{Z}$. Contrary to the usual situation in the infinite
geometry or geometries with
boundaries~\cite{finite-size-DH,general-DH} when only the $\phi=0$
stationary solution contributes, here all stationary solutions
contribute to the functional integral~(\ref{eq:functional-int}), since
when $\epsilon\to\infty$, all $X_{n}\to 0$.

Let us proceed more generally, supposing that the fugacities are
arbitrary and do not necessarily satisfy the pseudoneutrality
condition. Let $\psi_0\in\mathbb{R}$ be the solution of
\begin{equation}
  \label{eq:psi0}
  \sum_{\alpha}q_{\alpha}\zeta_{\alpha} e^{-\beta q_{\alpha}\psi_0}=0,  
\end{equation}
which is unique since the l.h.s.~of~(\ref{eq:psi0}) is a monotonous
function of $\psi_0$. Let us define $\zeta_{\alpha}^{*}=\zeta_{\alpha}
e^{-\beta q_{\alpha} \psi_0}$. These ``renormalized'' fugacities
satisfy the pseudoneutrality condition $\sum_{\alpha}
\zeta_{\alpha}^{*} q_{\alpha}=0$. Clearly the general solutions to
equation~(\ref{eq:minim-app}) are
\begin{equation}
  X_n=2\pi n \sqrt{\frac{2}{\beta e^2\epsilon}}+X_0\,,
    \quad n\in\mathbb{Z}
\end{equation}
with $X_0=-i\sqrt{(2\beta/\epsilon)}\psi_0$. Now, we proceed to apply
the steepest descent method to
evaluate~(\ref{eq:functional-int}). Notice that it is valid to apply
this method when $\beta e^2 \ll 1$ and $\zeta_{\alpha} R^2 \gg
1$. This last condition is necessary for the part of the integral
depending on $X$. Taking into account all the stationary points $X_n$
we have
\begin{eqnarray}
  \Xi_{\epsilon}& 
  =
  &
  \frac{1}{Z_0}
  \sum_{n\in\mathbb{Z}}
  \int \prod_{\ell\geq1,\,m}d\phi_{\ell m} dX
  e^{-\frac{\beta}{\epsilon}\left(\frac{2\pi n}{\beta
  e}-i\psi_0\right)^2+
  4\pi R^2\sum_{\alpha}\zeta_{\alpha}^{*}}
  \\
  &&\times
  \exp\left[
  -\frac{\beta}{4\pi}\sum_{\ell\geq 1,\,m}
  [\ell(\ell+1)+(\kappa R)^2] \mynormalordering{\phi_{\ell m}}\!\!^2
  -\frac{1}{2}(1+\epsilon (\kappa R)^2)(X-X_n)^2
  \right]
  \nonumber
\end{eqnarray}
with $\kappa=\sqrt{2\pi\beta\sum_{\alpha} q_{\alpha}^2
\zeta_{\alpha}^{*}}$ the inverse Debye length. Performing the
Gaussian integrals we find
\begin{equation}
\Xi_{\epsilon}=
\left[
\prod_{\ell\geq1,\,m} 
\left(
1+\frac{(\kappa R)^2}{\ell(\ell+1)}\right)
\prod_{k} e^{\kappa^2/\lambda_{k}^{0}}
\right]^{-1/2}
e^{V\sum_{\alpha}\zeta_{\alpha}^{*}}
f(\epsilon)
\end{equation}
where $\lambda_{k}^{0}=-\mathbf{K}^2$, $\mathbf{K}\in\mathbb{R}^2$,
are the eigenvalues of the Laplacian operator in the flat space
$\mathbb{R}^2$ (these appear from the self-energy term $v^{0}$ in the
hamiltonian, for details see~\cite{finite-size-DH,general-DH,3Dslab}),
and $V=4\pi R^2$ is the total area of the sphere. In the previous
equation we have defined
\begin{equation}
  f(\epsilon)=(1+(\kappa R)^2\epsilon)^{-1/2}
  \sum_{n\in\mathbb{Z}}
  e^{-\frac{\beta}{\epsilon}\left(\frac{2\pi}{\beta e}n-i\psi_0\right)^2}
\end{equation}
which can be expressed in terms of the Jacobi function
$\vartheta_3(u|\tau)=\sum_{n\in\mathbb{Z}} e^{i\pi\tau n^2+2n u i} $ as
\begin{equation}
  f(\epsilon)=(1+(\kappa R)^2\epsilon)^{-1/2}
  e^{\beta \psi_0^2/\epsilon}
  \vartheta_3\left(\left.\frac{2\pi\psi_0}{e \epsilon}\right|
  \frac{4\pi i}{\beta e^2 \epsilon}\right)
\end{equation}
To compute the limit of $\Xi_{\epsilon}$ when $\epsilon\to\infty$, it
is useful to use the Jacobi imaginary
transformation~\cite{Whittaker-Watson} to express $f(\epsilon)$ as
\begin{equation}
  f(\epsilon)=\sqrt{\frac{\beta e^2 \epsilon}{4\pi(1+(\kappa
  R)^2\epsilon)}}
  \ \vartheta_3\left(\left.\frac{\beta e \psi_0}{2}\right|
  \frac{i\beta e^2 \epsilon}{4\pi}\right)
  \,.
\end{equation}
The last term, the $\vartheta_3$ function, has limit 1 when
$\epsilon\to\infty$. Finally, taking $\epsilon\to\infty$ we obtain the
original partition function of the Coulomb system in the sphere
\begin{equation}
\label{eq:part-funct-sphere}
\Xi=
\left[
\frac{4\pi}{\beta e^2}
(\kappa R)^2
\prod_{\ell\geq1,\,m} 
\left(
1+\frac{(\kappa R)^2}{\ell(\ell+1)}\right)
\prod_{k} e^{\kappa^2/\lambda_{k}^{0}}
\right]^{-1/2}
e^{V\sum_{\alpha}\zeta_{\alpha}^{*}}
\,.
\end{equation}

It should be clear to the reader that the above calculations are very
general and can be easily adapted to other types of finite geometries
without boundaries. In that general case, the spherical harmonics and
$-\ell(\ell+1)/R^2$ are replaced, respectively, by the eigenfunctions
and the eigenvalues of the Laplacian in the considered geometry. For a
finite surface without boundaries, any constant function is an
eigenfunction of the Laplacian with zero eigenvalue. This constant
function plays the role of $Y_{00}$. Thus in an arbitrary finite
surface without boundaries the grand partition function of the Coulomb
gas is
\begin{equation}
\label{eq:part-funct-general}
\Xi=
\left[
\frac{\kappa^2 V}{\beta e^2}
\prod_{n, \lambda_n\neq 0} 
\left(
1-\frac{\kappa^2}{\lambda_n}\right)
\prod_{k} e^{\kappa^2/\lambda_{k}^{0}}
\right]^{-1/2}
e^{V\sum_{\alpha}\zeta_{\alpha}^{*}}
\,.
\end{equation}
where $\lambda_n$ are the eigenvalues of the Laplacian in the manifold
where the system lives and $V$ is the volume (area) of the manifold.
The result is similar to the one for a geometry with boundaries and
Dirichlet boundary conditions, found in Ref.~\cite{finite-size-DH},
except that it appears an additional term $\kappa^2 V/(\beta e^2)$
which is the contribution of the zero eigenvalue.

Notice that the role of the mean field $\psi_0$ is to renormalize the
fugacities, and the partition function is expressed in term of the
fugacities $\zeta_{\alpha}^*$ which satisfy the pseudoneutrality
condition. We recover the fact that the fugacities are not independent
controlling variables, this is due to the global neutrality of the
system, which we imposed from the start. The same situation arises in
geometries with boundaries, for details see the appendix B of
Ref.~\cite{finite-size-DH}.


\section{Thermodynamics of a Coulomb system on a sphere in the
  Debye--H\"uckel regime}
\label{sec:sphere-DH}

\subsection{Grand potential}

In this section we consider the sphere geometry and we compute
explicitly the partition function~(\ref{eq:part-funct-sphere}). The
results of this section are specific to the sphere.

The grand potential $\Omega=-k_B T\ln\Xi$ is given by
\begin{eqnarray}
  \beta \Omega &=&\frac{1}{2}\ln \frac{4\pi (\kappa R)^2}{\beta e^2}
  +\frac{1}{2}\ln \prod_{\ell=1}^{N}
  \left( 1+ \frac{R^{2}\kappa ^{2}}{\ell(\ell+1)}\right)^{2\ell+1}
  -(\kappa R)^{2}\int_{k_{\min }}^{K_{\max }}\frac{dK}{K}
  \nonumber\\
  &&
  -\sum_{\alpha} V \zeta_{\alpha }^{*}
  \label{granpotencial-sphereonshell}
\end{eqnarray}
The integral term in equation~(\ref{granpotencial-sphereonshell})
comes from the eigenvalues $\lambda_k^{0}$ of the Laplacian in the
flat space. The infinite product in
equation~(\ref{granpotencial-sphereonshell}) is divergent for large
$\ell$ and it should be cutoff to a maximum value $N$ of $\ell$. This
ultraviolet divergence is compensated by the one from the integral
when $K_{\max}\to\infty$. As explained and illustrated in several
examples in Refs.~\cite{finite-size-DH,general-DH}, the cutoffs
$K_{\max}$ and $N$ are proportional and the exact relation between
them can be found by requiring that in the limit $R\to\infty$ we
recover the bulk grand potential in the flat space which is known. In
the integral we also need to introduce an infrared cutoff
$k_{\min}$. It is explained in Refs.~\cite{finite-size-DH,general-DH}
that this cutoff is $k_{\min}=2e^{-C}/L$, where $C$ is the Euler
constant. The length $L$ is the same from equation~(\ref{eq:v0}) which
appears in the flat Coulomb potential $v^{0}$. In the Debye--H\"uckel
regime it is understood that $L$ is large, $L\to\infty$.

We can obtain an explicit expression for a certain regularization of
the infinite product appearing
in~(\ref{granpotencial-sphereonshell}). Let us consider the infinite
product
\begin{equation}
  \label{eq:prod-infini}
  P(z)=\prod_{\ell=1}^{\infty} \left(1+\frac{z^2}{\ell(\ell+1)}\right)^{\ell}
  e^{-z^2/\ell}
\end{equation}
which is convergent. Let us introduce the Barnes $G$
function~\cite{Barnes,Adamchik-Barnes}
\begin{equation}
  \label{eq:Barnes-G}
  G(z+1)=(2\pi)^{z/2} e^{-z(z+1)/2-Cz^2/2}
  \prod_{\ell=1}^{\infty} \left(1+\frac{z}{\ell}\right)^{\ell}
  e^{-z+z^2/(2\ell)}
\end{equation}
with $C$ the Euler constant. Let us write in
equation~(\ref{eq:prod-infini})
\begin{equation}
  1+\frac{z^2}{\ell(\ell+1)}=\frac{\displaystyle
    \left(1-\frac{\nu_1}{\ell}\right)
    \left(1-\frac{\nu_2}{\ell}\right)
    }{\displaystyle
    \left(1+\frac{1}{\ell}\right)}
\end{equation}
with 
\begin{equation}
  \nu_{1,2}=\left(-1\pm\sqrt{1-4z^2}\right)/2  
\end{equation}
that satisfy $\ell(\ell+1)+z^2=(\ell-\nu_1)(\ell-\nu_2)$. This allow
us to express the product $P(z)$ in terms of the Barnes $G$ function,
generalizing a standard procedure used to express infinite products in
terms of Gamma functions~\cite{Whittaker-Watson},
\begin{equation}
  P(z)=\prod_{\ell=1}^{\infty} \left(1+\frac{z^2}{\ell(\ell+1)}\right)^{\ell}
  e^{-z^2/\ell}
  =
  G(1-\nu_1)G(1-\nu_2)\,e^{-(1+C)z^2}
\end{equation}
Then we have
\begin{equation}
  \prod_{\ell=1}^{\infty}
  \left(
  1+\frac{z^2}{\ell(\ell+1)}\right)^{2\ell+1}
  e^{-2z^2/\ell}
  =\frac{[G(1-\nu_1)G(1-\nu_2)]^2
  e^{-2(1+C)z^2}}{\Gamma(1-\nu_1)\Gamma(1-\nu_2)}
\end{equation}
Using $\Gamma(\frac{1}{2}+x)\Gamma(\frac{1}{2}-x)=\pi/\cos(\pi x)$ and
$G(1+z)=\Gamma(z)G(z)$ we finally have
\begin{equation}
  \label{eq:infinite-product}
  \prod_{\ell=1}^{\infty}
  \left(
  1+\frac{z^2}{\ell(\ell+1)}\right)^{2\ell+1}
  e^{-2z^2/\ell}
  =\frac{\pi[G(-\nu_1)G(-\nu_2)]^2
  e^{-2(1+C)z^2}}{z^2\cosh([\pi\sqrt{4z^2-1}]/2)}
\end{equation}
Putting $z=\kappa R$ and replacing~(\ref{eq:infinite-product}) into
the expression~(\ref{granpotencial-sphereonshell}) for the grand
potential yields
\begin{eqnarray}
  \beta\Omega&=&\frac{1}{2}\ln \frac{4\pi (\kappa R)^2}{\beta e^2}
  +\frac{1}{2}\ln\frac{\pi[G(-\nu_1)G(-\nu_2)]^2}{(\kappa R)^2
    \cosh([\pi\sqrt{(2\kappa R)^2-1}]/2)}
  \\
  &&+(\kappa R)^2\left(-1-C+\sum_{\ell=1}^N \frac{1}{\ell}
  -\ln \frac{K_{\max}}{k_{\min}}
  \right)
  -V\sum_{\alpha} \zeta_{\alpha}
  \nonumber
\end{eqnarray}
In the limit where the ultraviolet cutoffs $K_{\max}$ and $N$ go to
infinity and replacing $k_{\min}=2e^{-C}/L$, we have
\begin{eqnarray}
  \beta\Omega&=&\frac{1}{2}\ln \frac{4\pi}{\beta e^2}
  +\frac{1}{2}\ln\frac{\pi[G(-\nu_1)G(-\nu_2)]^2}{\cosh(
    [\pi\sqrt{(2\kappa R)^2-1}]/2)}
  \\
  &&+(\kappa R)^2\left(-1+\ln \frac{N}{K_{\max}R}
    +\ln\frac{2e^{-C}R}{L}\right)
    -V\sum_{\alpha} \zeta_{\alpha}^{*}
    \,.
    \nonumber
\end{eqnarray}
We see that if the ultraviolet cutoffs $N$ and $K_{\max}$ are
proportional the expression for the grand potential is well
defined. Furthermore, in the limit $R\to\infty$ we should recover the
bulk value~\cite{finite-size-DH}
\begin{equation}
\frac{\beta \Omega_b }{V} =\frac{\kappa ^{2}}{4\pi }
\left[ -\ln \frac{\kappa
L}{2}-C +\frac{1}{2}\right] -\sum_{\alpha }\zeta_{\alpha }^{*}
\label{bulk2Dim}
\end{equation}
for the grand potential. We can compute this thermodynamic limit using
the known expansion of the Barnes $G$ function for large
argument~\cite{Barnes,Adamchik-Barnes,spectral functions}
\begin{equation}
  \label{eq:Barnes-Stirling}
  \ln G(1+z)\sim
  z^2\left(\frac{\ln z}{2}-\frac{3}{4}\right)+
  \frac{z}{2}\ln (2\pi)-\frac{\ln z}{12}+\zeta'(-1)+O(1/z)
  \,.
\end{equation}
($\zeta'(-1)$ is the derivative of the Riemann zeta function evaluated
at $-1$). Using this expansion we find that it is necessary that
$K_{\max}=N/R$ to recover the correct bulk grand potential in the
limit $R\to\infty$. Finally the grand potential for finite $R$ can be
expressed as
\begin{eqnarray}
  \beta\Omega&=&
  \frac{1}{2}\ln\frac{4\pi^2[G(-\nu_1)G(-\nu_2)]^2}{\beta e^2\cosh(
    [\pi\sqrt{(2\kappa R)^2-1}]/2)}
  +(\kappa R)^2\left(-1
    +\ln\frac{2e^{-C}R}{L}\right)
    \nonumber\\
    &&
    -V\sum_{\alpha} \zeta_{\alpha}^{*}
    \,.
    \label{eq:Omega-sphere-final-finite-R}
\end{eqnarray}
Using~(\ref{eq:Barnes-Stirling}), we can find the finite-size expansion
of the grand potential
\begin{equation}
  \beta\Omega=
  \beta\Omega_b+\frac{1}{3}\ln (\kappa R)
  +2\zeta'(-1)-\frac{1}{4}-\frac{1}{2}\ln\frac{\beta e^2}{4\pi}
  +O(1/(\kappa R))
\end{equation}
with the bulk grand potential $\Omega_b$ given by
equation~(\ref{bulk2Dim}). This expansion for the grand potential
shows that there is a finite-size logarithmic correction $(\chi/6)\ln
R$, where $\chi=2$ is the Euler characteristic of the sphere. Since
the system has no boundary there is no surface (perimeter) tension
term.


\subsection{Densities, pressure and the equation of state}

The density $n_{\alpha}$ of the species $\alpha$ can be obtained using
the standard thermodynamic relation $n_{\alpha}=\zeta_{\alpha}
\partial \ln \Xi/\partial \zeta_{\alpha}$. When computing this
derivative one has to take into account the fact that $\psi_0$ is a
function of $\zeta_{\alpha}$ implicitly given by
equation~(\ref{eq:psi0}). After some long but straightforward
calculation we finally find
\begin{eqnarray}
  \label{eq:densities}
  n_{\alpha}
  &=
  \zeta_{\alpha}^*\Bigg[1-&
    \left(\frac{\beta q_{\alpha}^2}{2}-\frac{\pi \beta^2 q_{\alpha}
      \sum_{\gamma}q_{\gamma}^3\zeta_{\gamma}^{*}}{\kappa^2}\right)
    \\
    &&\times
    \left(
    \psi(1+\nu_1)+\frac{\pi}{2}\cot(\pi\nu_1)+C+\ln\frac{L}{2R}
    \right)
    \Bigg] .
  \nonumber
\end{eqnarray}
We have used the relation~\cite{Adamchik-Barnes}
\begin{equation}
  \label{eq:derivative-G}
  \ln G(1+z)=\frac{z(1-z)}{2}+\frac{z}{2}\ln(2\pi)+\int_0^z
  x\psi(x)\,dx
\end{equation}
where $\psi(x)=d\ln\Gamma(x)/dx$ is the psi function. This relation
allows us to express the derivative of the Barnes $G$ function in
terms of the psi function $\psi(x)$. Also we used some known
identities~\cite{grad} satisfied by the psi function, such as
$\psi(\nu)+\psi(-\nu-1)=2\psi(\nu+1)-[\nu(\nu+1)]^{-1}+\pi\cot(\pi\nu)$.
One can easily verify that the system is neutral $\sum_{\alpha}
q_{\alpha} n_{\alpha}=0$. The total density $n=\sum_{\alpha}
n_{\alpha}$ is
\begin{equation}
  n=\sum_{\alpha} \zeta_{\alpha}^{*}
  -\frac{\kappa^2}{4\pi} \left(
    \psi(1+\nu_1)+\frac{\pi}{2}\cot(\pi\nu_1)+C+\ln\frac{L}{2R}
    \right)
    \,.
\end{equation}
The finite-size expansion of the densities, when $R\to\infty$, reads
\begin{equation}
  n_{\alpha}=n_{\alpha}^b
  -\frac{\zeta_{\alpha}^b}{6(\kappa R)^2}
    \left(\frac{\beta q_{\alpha}^2}{2}-\frac{\pi \beta^2 q_{\alpha}
      \sum_{\gamma}q_{\gamma}^3\zeta_{\gamma}^{*}}{\kappa^2}\right)
\end{equation}
where the bulk density is
\begin{equation}
  n_{\alpha}^b=\zeta_{\alpha}^{*}
  \left(
  1-\left(\frac{\beta q_{\alpha}^2}{2}-\frac{\pi \beta^2 q_{\alpha}
      \sum_{\gamma}q_{\gamma}^3\zeta_{\gamma}^{*}}{\kappa^2}\right)
  \ln\frac{\kappa L e^C}{2}
  \right)
  \,.
\end{equation}
And for the total density
\begin{equation}
  n=n^b+\frac{1}{6}\frac{1}{4\pi R^2}  
\end{equation}
with $n^b=\sum_{\alpha} n_{\alpha}^b$. Thus the total number of
particles $N=N_b + 1/6$, where $N_b$ is the bulk number of
particles. The $1/6$ finite-size correction to the number of particles
is a consequence of the $(1/3) \ln R$ correction to the grand
potential.

The pressure $p=-\partial\Omega/\partial V$ is given by
\begin{equation}
  \beta p=\sum_{\alpha} \zeta_{\alpha}^{*}
  -\frac{\kappa^2}{8\pi}
  +\frac{\kappa^2}{4\pi}
  \left(\psi(1+\nu_1)+\frac{\pi}{2}\cot(\pi\nu_1)
  +\ln \frac{L e^{C}}{2R}\right)
  \,.
\end{equation}
Using equation~(\ref{eq:densities}), and neglecting terms of higher
order than $\beta e^2$, we find the equation of state
\begin{equation}
  \beta p=\sum_{\alpha} n_{\alpha}\left(1-\frac{\beta q_{\alpha}^2}{4}
  \right)
  \,.
\end{equation}
We remind the reader that, from a scale invariance analysis, one can
show that this equation of state is actually valid in the whole range
of stability of the system of point particles, both in the flat
geometry and in the sphere.

\subsection{Internal energy}

The excess internal energy is $U_{\exc}=(\partial (\beta
\Omega)/\partial \beta)_{\zeta_{\alpha},
V}$. Using~(\ref{eq:Omega-sphere-final-finite-R}), we find
\begin{equation}
  \label{eq:Uexc}
  \beta U_{\exc}=
  -(\kappa R)^2 \left(
  \psi(1+\nu_1)+\frac{\pi}{2}\cot(\pi\nu_1)+C+\ln\frac{L}{2R}
  \right) -\frac{1}{2}
\end{equation}
We recall that $\nu_1=\left(-1+\sqrt{1-(2\kappa R)^2}\right)/2$. 

We can derive this result~(\ref{eq:Uexc}) using a more traditional
approach to the Debye--H\"uckel theory. Consider a unit pseudocharge
(charge plus neutralizing background) located at the north
pole. Following the usual formulation of Debye--H\"uckel theory, this
pseudocharge is screened by a polarization cloud created by the
plasma. This cloud has a charge density given by
\begin{eqnarray}
  \rho_{\mathrm{pol}}(\theta,\varphi)&=& \sum_{\alpha} q_{\alpha}
  n_{\alpha} e^{-\beta q_{\alpha} K(\theta,\varphi;0,0)} - \int
  \sum_{\alpha} q_{\alpha} n_{\alpha} e^{-\beta q_{\alpha}
  K(\theta',\varphi';0,0)}\ \frac{d\anglesol'}{4\pi} \nonumber\\
  \label{eq:DH-density}
  &\simeq&
  -\kappa_{D}^2 K(\theta,\varphi;0,0)
  +\kappa^2_{D} \langle K \rangle
\end{eqnarray}
where $K(\theta,\varphi;0,0)$ is the (mean field) electric potential
created at $(\theta,\varphi)$ by the pseudocharge and its polarization
cloud.  Also $\langle K \rangle=\int
K(\theta',\varphi';0,0)\,d\anglesol'/(4\pi)$ is the average of
$K$ over the sphere. To understand the second term
in~(\ref{eq:DH-density}) recall that we are dealing with
pseudocharges (charges plus neutralizing background), and the
polarization cloud is also made of pseudocharges. In the second line
we linearized the exponentials as it is usually done in the
Debye--H\"uckel theory and defined the inverse Debye length
$\kappa_{D}=\sqrt{2\pi\beta\sum_{\alpha} n_{\alpha}
q^{2}_{\alpha}}$. At the Debye--H\"uckel level of approximation
$\kappa_{D}\simeq\kappa$. The potential $K$ satisfies the modified
Poisson equation
\begin{equation}
  \Delta K = -2\pi\left(
  \delta-\frac{1}{4\pi R^2}+\rho_{\mathrm{pol}}(\theta,\varphi)
  \right)
\end{equation}
and using~(\ref{eq:DH-density}) we arrive at the modified
Debye--H\"uckel equation for the sphere geometry
\begin{equation}
  \label{eq:DH-equation-sphere}
  \Delta K -\kappa^2 K + \kappa^2 \langle K \rangle 
  =
  -2\pi\left(\delta-\frac{1}{4\pi R^2}\right)
\end{equation}
As in the case of the modified Poisson
equation~(\ref{eq:Poisson-sphere}), we can notice that the solution of
the modified Debye--H\"uckel equation~(\ref{eq:DH-equation-sphere}) is
determined up to an arbitrary additive constant. Indeed this can be
clearly seen if we look for a solution of
equation~(\ref{eq:DH-equation-sphere}) as an expansion in spherical
harmonics
\begin{eqnarray}
  \label{eq:DH-harmonics}
  K(\theta,\varphi;0,0)&=&\sum_{\ell=1}^{\infty} \sum_{m=-\ell}^{\ell}
  \frac{2\pi}{\ell(\ell+1)+(\kappa R)^2} 
  \overline{Y_{\ell m}(0,0)} Y_{\ell m}(\theta,\varphi)
  \\
  &&+ b_0 \overline{Y_{00}(0,0)} Y_{00}(\theta,\varphi)
  \,.
  \nonumber
\end{eqnarray}
The constant term $b_0$, for $l=0$, cannot be determined from the
differential equation~(\ref{eq:DH-equation-sphere}). Thus the average
$\langle K \rangle=b_{0}/(4\pi)$ can be chosen arbitrarily. However
as we will see below this constant term will not be needed in the
following.

The excess internal energy $U_{\mathrm{exc}}$ can be computed from the
Debye--H\"uckel potential $K$ as the potential energy of the
pseudocharge in the north pole, in the potential field $K$
\begin{eqnarray}
  \label{eq:U-DH-1}
  U_{\mathrm{exc}}&=\frac{4\pi R^2}{2} &
  \sum_{\alpha}q_{\alpha}^2
  n_{\alpha} 
  \\
  &&
  \!\!\!\!
  \lim_{\atop{\theta\to 0}{r=R\theta\to 0}}
  \left[
  \int K(\theta,\varphi;0,0)
  (\delta(\theta,\varphi;0,0)-1/(4\pi R^2))R^2d\anglesol
  +\ln\frac{r}{L} \right]
  \nonumber
\end{eqnarray}
Notice that we subtract the self-energy $-\ln (r/L)$ corresponding to
a flat geometry, in accordance to the
prescription~(\ref{eq:prescript-self-s}), to obtain a finite
result. This kind of prescription has also been used in
Ref.~\cite{Janco-pressures-Coulomb} for the derivation of the pressure
of the Coulomb gas using the Maxwell stress tensor. We have
\begin{equation}
  \label{eq:U-DH}
  U_{\mathrm{exc}}=\frac{4\pi R^2}{2}  
  \sum_{\alpha}q_{\alpha}^2 n_{\alpha} 
  \lim_{\atop{\theta\to 0}{r=R\theta\to 0}}
  \left(
  K(\theta,\varphi;0,0)-\langle K \rangle+\ln\frac{r}{L}
  \right)
\end{equation}
Thus we only need $K-\langle K \rangle$, the constant term $\langle K
\rangle$ is irrelevant for the calculation of the internal energy. An
explicit expression for $K$ can be obtained by noticing that the
Yukawa potential $Y$, which is the solution of
\begin{equation}
  \label{eq:Yukawa-sphere}
  \Delta Y - \kappa^2 Y =-2\pi\delta
  \,,
\end{equation}
is also a particular solution of
equation~(\ref{eq:DH-equation-sphere}) with $\langle Y
\rangle=1/(2\kappa^2 R^2)$. 
Therefore, $Y$ has an expansion in spherical harmonics of the
form~(\ref{eq:DH-harmonics}) with $b_0=2\pi/(\kappa^2 R^2)$. So, the
difference between $K$ and $Y$ is a constant equal to $-1/[2(\kappa
R)^2]+ \langle K \rangle$. On the other hand
equation~(\ref{eq:Yukawa-sphere}) can be solved directly since it reduces
to a Legendre equation in the variable $\cos\theta$. The solution
is~\cite{Janco-pressures-Yukawa}
\begin{equation}
  Y(\theta)=-\frac{\pi}{2\sin(\nu_1 \pi)}
  P_{\nu_1}(-\cos\theta)
\end{equation}
with $P_{\nu_1}$ a Legendre function. Finally we find the
Debye--H\"uckel potential for a pseudocharge located at the north
pole
\begin{equation}
  K(\theta,\varphi;0,0)-\langle K \rangle=-\frac{\pi}{2\sin(\nu_1
  \pi)} P_{\nu_1}(-\cos\theta) -\frac{1}{2(\kappa R)^2}
  \,.
\end{equation}
As $\theta\to 0$, the Legendre function has the
behavior~\cite{Janco-pressures-Yukawa}
\begin{equation}
  P_{\nu_1}(-\cos\theta)=
  \frac{2\sin(\nu_1\pi)}{\pi}
  \left[
    \ln\sin\frac{\theta}{2}+C+\psi(1+\nu_1)
    +\frac{\pi}{2}\cot(\nu_1\pi)
    \right]
  +o(1)
  \,.
\end{equation}
Using this asymptotic behavior into equation~(\ref{eq:U-DH}) allow us
to retrieve the result~(\ref{eq:Uexc}) for the internal energy.


\section{Finite-size corrections for Coulomb systems on finite surfaces
  without boundaries}
\label{sec:finite-size}

In this section we consider a Coulomb gas confined in an arbitrary
finite surface without boundaries. We will compute the finite-size
expansion of the grand potential. Let $R$ be the square root of the
total area $V$ of the surface. Following~\cite{general-DH}, we
introduce the zeta function of the Laplacian in the geometry
considered,
\begin{equation}
  \label{eq:zeta-fnct}
  Z(s, a)=\sum_{k=1}^{\infty} (a-\lambda_{k})^{-s}
\end{equation}
where $\lambda_{k}$ are the eigenvalues of the Laplacian. Notice that
we omit the vanishing eigenvalue $\lambda_0=0$ in the
definition~(\ref{eq:zeta-fnct}).

In Ref.~\cite{general-DH}, this zeta function is related to grand
potential. We can directly transpose the calculations of
Ref.~\cite{general-DH} to the present case, taking special care of the
additional contribution to the grand potential due to the vanishing
eigenvalue. Then, we obtain for the grand potential the relation
\begin{equation}
  \label{eq:omega-zeta}
  \beta \Omega=
  \frac{1}{2}\left[
    Z'(0,0)-Z'(0,\kappa^2)
    \right]
  +\frac{\kappa^2 V}{4\pi}\ln\frac{2e^{-C}}{L}
  +\frac{1}{2}\ln\frac{\kappa^2 V}{\beta e^2}
  -V\sum_{\alpha}\zeta_{\alpha}^*
  \,.
\end{equation}
The prime means differentiation with respect to first variable of the
zeta function. The finite-size expansion of the zeta functions
involved in~(\ref{eq:omega-zeta}) can be obtained from the known
small-argument expansion of the heat kernel,
$\Theta(t)=\sum_{n=0}^{\infty} e^{t\lambda_{k}}$, of the
Laplacian~\cite{Kac,Singer} which reads,
\begin{equation}
  \Theta (t)=\frac{V}{4\pi t} +\frac{\chi}{6} +o(t^{1/2}) 
  \label{expansion-HK-2D}
\end{equation}
with $\chi$ the Euler characteristic of the surface. This is explained
in detail in Ref.~\cite{general-DH}. Here we only need to take special
care of the fact that we omitted the zero eigenvalue in the zeta
function, which is equivalent to subtract $1$ to the heat
kernel. Thus, following~\cite{general-DH}, we find
\begin{equation}
   \label{eq:zeta-reg-gp*-2D} 
   \frac{1}{2}\left[
     Z'(0,0)-Z'(0,\kappa^2)
     \right]
   = \frac{  \kappa^2 }{4\pi}
   \left(\frac{1}{2}-\ln \kappa\right) V
   +\left[\frac{\chi}{6}-1\right]\ln(\kappa R)
   +O(1)
\end{equation}
Replacing into~(\ref{eq:omega-zeta}) we finally find the finite-size
expansion of the grand potential
\begin{equation}
  \label{eq:finite-size-exp-2D}
  \beta \Omega = \beta \Omega_b +\frac{\chi}{6}\ln(\kappa R) +O(1)
\end{equation}
with the bulk grand potential $\Omega_b$ given by~(\ref{bulk2Dim}).
Notice the existence of a logarithmic finite-size correction $(\chi/6)
\ln R$, which is universal, i.~e.~independent of details of the
microscopic constitution of the system. Actually this universal
finite-size correction seems to exist even beyond the low-coupling
regime considered here. In particular it has been shown to exist for
the sphere geometry for the one-component
plasma~\cite{Janco-Trizac-sphere-ocp} and the two-component plasma
both in its charge symmetric~\cite{JancoKalinaySamaj-sphere-tcp} and
charge asymmetric version~\cite{Samaj-sphere-tcp-asym}, in the whole
range of stability of the system of point particles.

\section{Summary and perspectives}

We have showed how to build the Debye--H\"uckel theory for
two-dimensional Coulomb systems confined in a finite surface without
boundary. In particular we showed how to perform the sine-Gordon
transformation for this kind geometry and how to map the statistical
mechanics problem into a field theory on the finite surface. This
could have further applications, for instance, to compute higher order
corrections in $\beta e^2$ to the grand potential and other
thermodynamic quantities.  

For the case of the sphere geometry we explicitly computed the grand
potential and other thermodynamic quantities. For a general geometry,
we showed the existence of a universal logarithmic finite-size
correction for the grand potential.


\ack 
Partial financial support from COLCIENCIAS project 1204-05-13625 and
ECOS-Nord/COLCIENCIAS-ICFES-ICETEX is acknowledged. I would like to
thank B.~Jancovici for several enlightening discussions.


\newcommand{\bio}[5]{#1, \textit{#2} \textbf{#3} (#5) #4.} 



\end{document}